# Integrating Competency-Based Education in Interactive Learning Systems

**Maximilian Sölch, Moritz Aberle, Stephan Krusche**
Technical University Munich
maximilian.soelch@tum.de, moritz.aberle@tum.de, krusche@tum.de

**ABSTRACT**: Artemis is an interactive learning system that organizes courses, hosts lecture content and interactive exercises, conducts exams, and creates automatic assessments with individual feedback. Research shows that students have unique capabilities, previous experiences, and expectations. However, the course content on current learning systems, including Artemis, is not tailored to a student's competencies. The main goal of this paper is to describe how to make Artemis capable of competency-based education and provide individual course content based on the unique characteristics of every student. We show how instructors can define relations between competencies to create a competency relation graph, how Artemis measures and visualizes the student's progress toward mastering a competency, and how the progress can generate a personalized learning path for students that recommends relevant learning resources. Finally, we present the results of a user study regarding the usability of the newly designed competency visualization and give an outlook on possible improvements and future visions.

## 1    INTRODUCTION

A successful and individual learning experience requires that students receive formative feedback and insights into their learning progress in a course (Scheffel et al., 2014). Students should be able to adapt the course content based on their competencies and needs. Instructors want to know whether their students gained specific competencies while attending their course. However, such functionality is unavailable or works poorly in existing learning management systems.

Artemis allows instructors to create a scalable and productive student-instructor setting, even for extensive or remote-only courses. The platform aspires to reduce the overall workload for instructors with large audiences without compromising the benefits of an interactive learning environment (Krusche et al., 2017; Krusche & Seitz, 2018). It focuses on active student participation using automated or manual software-assisted grading and iterative feedback. In addition, Artemis supports lecture units, such as video, text, file, online, and exercise units. More than 10,000 students from several universities across Germany and Austria actively use Artemis (Krusche, 2021).

## 2    APPROACH

Instructors should be able to teach their courses using a fundamentally different approach. With competency-based education, they can outline the desired student abilities at the end of the course and then provide content to help students reach these learning objectives (Curry & Docherty, 2017). A learning system must track students' progress and allow instructors to define relations between different competencies. Instructors can create a graph network of learning objectives by marking a competency as a prerequisite or adding more detailed sub-topics to a competency. These dependencies between competencies allow students to find a suitable and personalized learning path throughout the defined competencies to progress in the course individually. In addition, the learning system needs to visualize the progress toward mastering a competency for the students and the course instructors.

## 2.1 Competency Relations

To allow instructors to define relations between different competencies, we designed and implemented the user interface shown in Figure 1. Instructors can create directed relations between exactly two competencies of the same course by specifying the relation's head and tail competency as well as the relation's type. A relation can have one of the following types: *assumes*, because a competency can require that another competency is mastered first, *extends*, because a competency can add new aspects to another competency, *relates*, because a competency can be connected to another competency, and *matches*, because a competency can be identical to another competency. A relation cannot be reflexive and no more than one relation of each type can exist between two competencies. To get a better overview of all the existing relations between a course's competencies, Artemis generates a competency relation graph. This graph updates when an instructor adds or removes a relation.

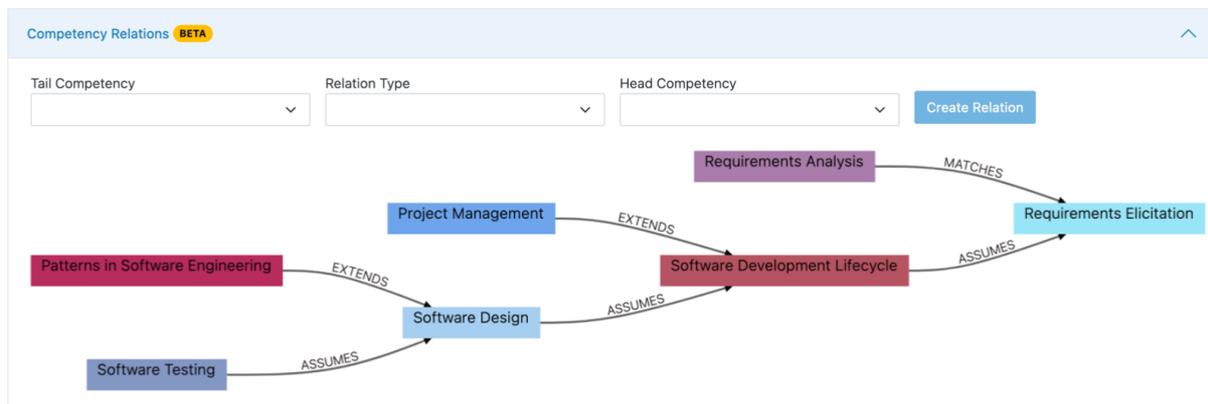

Figure 1: Defining relations between competencies in Artemis using a competency graph

## 2.2 Learning Analytics

To measure the student's progress in a course, Artemis tracks the active participation in exercise units and uses a checkbox for visualizing the completion status of a lecture unit. Besides the visual indication, the checkbox functions as a button for the user to manually toggle the completion status at the same time. Artemis also marks lecture units as completed automatically depending on their type when the following conditions are met: Once the user clicks the button to download an attachment file, the corresponding file unit is considered complete. Text units are completed as soon as the student clicks on the unit to open the collapsed text. Students automatically complete an online unit when they click on the link forwarding them to the external website. Video units are marked as completed five minutes after clicking on the unit to reveal the embedded video. In addition to the completion status, Artemis also provides different exercise statistics for students to track their performance and compare themselves to the course average and for instructors to get insights into their course's performance.

### 2.3 Different Metrics for Competencies

Artemis calculates different metrics and displays them to the students allowing them to get a quick overview of their learning progress regarding a competency. They can track their mastery advancement for each competency by viewing the progress value, calculated from the number of completed learning resources linked to the competency, and the confidence value, composed of the average score for all exercises linked to the competency. A competency is considered mastered if the progress is at 100% and the confidence value is equal to or higher than the defined mastery threshold of the competency. Inspired by the Apple Watch's fitness rings, Artemis shows the user three individual progress bars in a circle depicted in Figure 2.

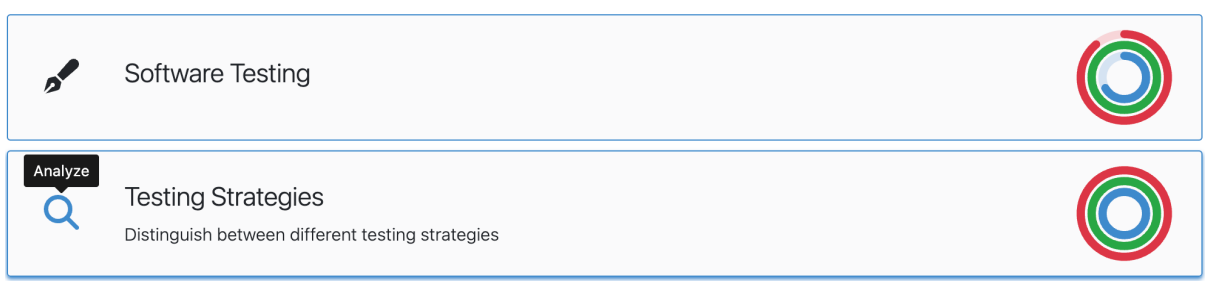

Figure 2: Visualization of competencies and a student's progress in Artemis

The innermost blue ring visualizes the overall progress P, which is the percentage of completed lecture units and participated exercises linked to the competency. The second, green progress bar shows the confidence level C of the competency in proportion to the threshold value T (mathematically $C * \frac{1}{T}$). As an example, if the student's latest confidence level equals 60% and the mastery threshold is set to 80%, the ring would be 75% full. The outermost red ring represents the advancement toward mastery as a weighted function (with $w = \frac{2}{3}$) of progress and confidence defined as $(1 - w) * P + w * C * \frac{1}{T}$. If a student has mastered the competency without completing all linked learning resources, we override the progress of the mastery ring to 100% nevertheless. Competencies can be categorized according to Bloom's revised taxonomy (Krathwohl, 2002). In the visualization we use different icons that support a quick identification of a competency according to the taxonomy. Instructors can link competencies to learning resources as prerequisites or as learning goals. In combination with the competency relation graph and the student's competency mastery, Artemis uses these links to recommend suitable learning resources and generate a personalized learning path for students.

### 3 EVALUATION

To evaluate the newly designed visualization of competencies and a student's progress in Artemis, we conducted the short version of the User Experience Questionnaire (UEQ-S) with 7 participants (Schrepp et al., 2017). All of them are students randomly invited from a TUM university course and are familiar with Artemis. During the UEQ-S the participants rate pairs of terms with opposite meanings on a 7-point Likert scale and their answers are scaled from -3 to +3.

The results of the UEQ-S attest the proposed visualization a good usability overall with a mean of 1.37. In particular, the results show an excellent hedonic quality with a mean value of 2.06 indicating that the user interface is appealing and not boring. The pragmatic quality, which relates to the practicality and functionality of a user interface, is slightly lower with a mean value of 0.714 and shows potential for additional improvements. One possible improvement could be to add a legend explaining the meaning of the rings as well as the icons. During the evaluation, most of the participants mentioned that this improvement would have made the visualization clearer and easier to understand. Therefore, we plan to add an informative legend to the visualization before we release it to the public.

## 4    CONCLUSION

We presented an approach to integrating competency-based education into learning management systems, specifically for Artemis, including a new and innovative visualization of a student's progress toward mastery of a competency. To evaluate this new visualization, we conducted a UEQ-S that showed promising results for the user experience.

Future work includes improving the calculation of metrics for competencies. For example, mastery of competencies might slowly decline after the final exam. Artemis could account for this fact and model knowledge decay within the adaptive learning system (Bergeron, 2014). Another potential area of improvement is the visualization of learning paths for students and instructors. Artemis uses a graphics-based interface to show an instructor the relationships between all competencies in the proposed system. However, it is difficult for the user to derive all possible learning paths from this representation. Artemis does not yet represent the learning paths chosen by students in the course. Visualizing how certain groups are progressing in the competencies would be a fascinating insight for the instructor in the context of learning analytics.